\input phyzzx

\def\IR{{\hbox{{\rm I}\kern-.2em\hbox{\rm R}}}}
\def\IB{{\hbox{{\rm I}\kern-.2em\hbox{\rm B}}}}
\def\IN{{\hbox{{\rm I}\kern-.2em\hbox{\rm N}}}}
\def\IC{{\ \hbox{{\rm I}\kern-.6em\hbox{\bf C}}}}

\def\IZ{{\hbox{{\rm Z}\kern-.4em\hbox{\rm Z}}}}
\def\to{\rightarrow}

\def\underarrow#1{\vbox{\ialign{##\crcr$\hfil\displaystyle
{#1}\hfil$\crcr\noalign{\kern1pt
\nointerlineskip}$\longrightarrow$\crcr}}}
%

\def\ltorder{\mathrel{\raise.3ex\hbox{$<$}\mkern-14mu
             \lower0.6ex\hbox{$\sim$}}}
\def\lesssim{\mathrel{\raise.3ex\hbox{$<$}\mkern-14mu
             \lower0.6ex\hbox{$\sim$}}}


\font\Ilis=cmbx10 scaled \magstep2

\vbox{\vskip 2cm}
\input phyzzx
\overfullrule=0pt
\tolerance=5000
\overfullrule=0pt
\twelvepoint

\twelvepoint

\titlepage

\title{\bf GEOMETRY ASSOCIATED WITH SELF-DUAL YANG-MILLS AND THE CHIRAL 
MODEL APPROACHES TO SELF-DUAL GRAVITY} 
\author{Hugo Garc\'{\i}a-Compe\'an$^a$, \foot{E-mail: 
compean@fis.cinvestav.mx} Jerzy F. Pleba\'nski$^a$ \foot{E-mail: 
pleban@fis.cinvestav.mx} and Maciej Przanowski$^b$ } 

\medskip

\address{ $^a$Departamento de F\'{\i}sica 
\break  Centro de Investigaci\'on y de Estudios Avanzados del IPN.
\break Apdo. Postal 14-740, 07000, M\'exico D.F., M\'exico.} 
\vskip .5truecm
\centerline{\it $^b$Institute of Physics}
\centerline{\it Technical University of \L \'od\'z}
\centerline{\it W\'olcza\'nska 219,  93-005 \L \'od\'z, Poland.}


\abstract{A geometric formulation of the Moyal deformation for the
Self-dual Yang-Mills theory and the Chiral Model approach to
Self-dual gravity is given. We find in Fedosov's geometrical construction
of deformation quantization the natural geometrical framework associated
to the Moyal deformation of the six-dimensional version of the second
heavenly equation and the Park-Husain heavenly equation. The
Wess-Zumino-Witten-like Lagrangian of Self-dual gravity is re-examined
within this context.}


\endpage

\chapter{\bf Introduction}

The purpose of this paper is to describe some conjectures in the geometry
of deformation quantization for Self-dual Yang-Mills (SDYM) theory and
the Chiral Model approach to Self-dual Gravity (SDG). This relation was
originally suggested by I.A.B. Strachan in [1]. He has developed a
deformed differential commutative geometry and has applied it to describe, within this
geometrical framework, the multidimensional integrable systems. Here we intend to consider the application of some {\it non-commutative} geometry (Fedosov's geometry) to self-dual gravity.

The relation, for instance, between SDYM theory, Conformal Field Theory
and Principal Chiral Model, all them with gauge group SDiff $(\Sigma)$
(area-preserving diffeomorphism group of two-dimensional simply connected
and symplectic manifold $\Sigma$), has been quite studied only at the
algebraic level [2-10]. The standard approach consists in studyng a
classical field theory invariant under some symmetry group, for instance,
SU$(N)$. In the case of full Yang-Mills theory its large-$N$ limit ($N \to
\infty$) is somewhat mysterious, however it is very necessary to understand it in the searching for new faces of integrability [11].  Drastic
simplifications in some classical equations seem to confirm these
speculations [12]. However, geometric and topological aspects of the
correspondence SDYM and SDG remain to be clarified [13].

Integrable deformations of SDG, in particular the Moyal deformation of the
first heavenly equation were studied by Strachan [14] and K. Takasaki [15].
Moyal deformation of the second heavenly equation was considered in
[16,17]. The Weyl-Wigner-Moyal (WWM) formalism has been very useful in
order to find (beginning from SDYM theory) a version of the `master
system' (compare with [2]) which leads to different versions of the
heavenly equations in SDG [18,19]. The application of WWM-formalism to
the Lagrangians and equations of motion of SDG for a
sdiff$(\Sigma)$-valued scalar field on ${\cal M}$, leads to the Moyal
deformation of the six-dimensional version of the second heavenly equation
(sdiff$(\Sigma)$ can be seen as the Lie algebra of SDiff$(\Sigma)$). This equation
and its associated Lagrangian was found in [20]. The application of the
WWM-formalism to the Principal Chiral Model approach to dynamical SDG was
done in terms of a scalar field by us in Ref. [21]. This dynamics was
enclosed in the P-H heavenly equation [21]. In [21] also some
explicit solutions were constructed using an explicit Lie algebra
representation of linear operators acting on some Hilbert space. As an
intermediate step of constructing solutions in SDG using WWM-formalism one
can represent the dynamics using a Moyal deformation of the P-H heavenly
equation [21]. WWM-formalism also has been very useful to show that some
results about harmonic maps [22] can be carried over to SDG [23]. As a
consequence one can define the ``Gravitational Uniton'' which seems to
be simultaneously a uniton {\it i.e.}, an appropriate solution of chiral
equations [22] and a solution of the P-H heavenly equation describing the 
SDG metric.  A WZW-like action for SDG can be constructed as well. ``How
can one to interpret geometrically this WZW action and the associated
``Wess-Zumino'' term''? was a question proposed in the last part of [23]. 
Here we intend to give an answer to this question.  In this paper, in a
spirit of Strachan [1], we describe some conjectures about the
geometry associated to integrable deformations of SDG and its
possible relation with the standard geometry of gauge theory.  The paper
is organized as follows: To be the most self-contained, in
Section 2 we briefly review the geometrical construction of deformation
quantization given by Fedosov [24]. In Section 3 we discuss the Moyal
deformation of SDYM theory, the associated six-dimensional version of the
second heavenly equation [20] and P-H heavely equation given in Ref. [21]. 
In Section 4 we describe within Fedosov's geometrical construction, the
Moyal deformation of the SDYM equations via Yang's and
Donaldson-Nair-Schiff equations. In particular, the Moyal deformation of
the six-dimensional version of the second heavenly equation is considered. In this same
section we discuss the geometry associated with the principal chiral model
approach to SDG. The Moyal deformation of the P-H heavens is described as
well.  Section 5 is devoted to study the WZW-like Lagrangian found in [23]
in the same geometrical framework. Finally in section 6 we give our final
remarks. 

\vskip 2truecm


\chapter{ \bf Fedosov's Geometry of Deformation Quantization}

Deformations of the Poisson Lie algebra structure on symplectic manifolds
have been studied by many authors [25]. Recently some extensions of the
WWM-formalism have been carried over to the phase space as represented by
cotangent bundle [26]. More recently interesting connections with
non-commutative geometry have been studied by M. Reuter in [27].

In this section we review some aspects of the geometry of deformation
quantization given by Fedosov in Ref. [24].  The most important objects we
consider here involve Weyl algebra bundle, differential forms and the
trace formula defined for this algebra.

\vskip 1truecm 

\section{ Weyl algebra bundle}

Let $({\cal M}, \omega)$ be a symplectic manifold of dimension $2n$ and
$\omega$ the corresponding symplectic form on ${\cal M}$. The formal Weyl
algebra ${\cal W}_x$ associated with the tangent space at the point $x \in
{\cal M}, \ \ T_x {\cal M},$ is the {\it associative} algebra over {\bf C}
with a unit. An element of ${\cal W}_x$ can be expressed by

 $$a(y) = \sum_{2k +l \geq 0} \hbar^k a_{k, i_1 \ldots i_l} y^{i_1} 
\ldots y^{i_l}, \eqno(2.1)$$
where $\hbar$ is the deformation parameter, $y = (y^1, \ldots ,y^{2n}) \in
T_x {\cal M}$ is a tangent vector and the coefficients $a_{k,i_1 \ldots 
i_l}$ constitute the symmetric covariant tensor of degree $l$ at $x \in
{\cal M}$.  

The product on ${\cal W}_x $ which determines the associative algebra
structure is defined by

$$ a \bullet b \equiv {\rm exp} \bigg( +{i\hbar\over 2} \omega^{ij} 
{\partial \over 
\partial y^i}{\partial \over \partial z^i}\bigg) a (y ,\hbar) b (z, 
\hbar) |_{z=y} $$

$$= \sum^{\infty}_{k=0} (+ {i\hbar\over 2})^k {1 \over k!} \omega^{i_1
j_1} \ldots \omega^{i_kj_k} {\partial^k a\over \partial y^{i_1} \ldots
\partial y^{i_k}} {\partial ^k b \over \partial y^{j_1} \ldots \partial
y^{j_k}}, \eqno(2.2)$$ for all $a, b \in {\cal W}_x$. Here $\omega^{ij}$
are the components of the tensor inverse to $\omega_{ij}$ at $x$. Of
course the product ``$\bullet$'' is independent of the basis. 

Having this one can define an algebra bundle structure taking the disjoint
union of Weyl algebras for all points $x \in {\cal M} \ i.e. \ \tilde
{\cal W} = {\parallel \over x \in M} {\cal W}_x$. $\tilde{\cal W}$ is the
total space and the fiber is isomorphic to a Weyl algebra ${\cal W}_x$. 
Thus we have the Weyl algebra bundle structure

$$ \tilde{\cal W} \buildrel{\pi}\over{\to} {\cal M}, \ \ \ \ \ {\cal W}_x
\cong \pi^{-1}(\{x\}),
\eqno(2.3)$$ where $\pi$ is the canonical projection.

Let ${\cal E}( \tilde {\cal W})$ be the set of sections of $\tilde {\cal
W}$ which also has a Weyl algebra structure with unit. Denote by $a(x, y,
\hbar)$ an element of ${\cal E} (\tilde {\cal W})$; it can be written as
follows

$$ a (x, y, \hbar) = \sum_{2k + l \geq 0} \hbar ^k \ a_{k, i_1 \ldots i_l}
(x) y^{i_1} \ldots y^{i_l}, \eqno(2.4) $$ where $y = (y^1,...,y^{2n}) \in
T_x{\cal M}$ is a tangent vector, $a_{k, i_1 \ldots i_l}$ are smooth
functions on ${\cal M}$ and $x \in {\cal M}$. 

\vskip 1truecm
\section{Differential Forms}

In Section 4 we shall define a field theory on space-time with the fields
taking values in the Weyl algebra of sections ${\cal E}(\tilde {\cal W})$
(instead of the usual Lie algebra). In order to do that we need the notion
of $ \tilde{\cal W}$-valued differential $q$-form on ${\cal M}$. A
$q$-form can be written as

$$ a = \sum  \hbar^k a_{k, j_1...j_q}(x,y) dx^{j_1}\wedge ... \wedge 
dx^{j_q}$$

$$ = \sum_{2k + p \geq 0} \hbar ^k a_{k, i_1 \ldots i_p, j_1 \ldots j_q}
(x)  y^{i_1} \ldots y^{i_p} d x^{j_1}\wedge \cdots \wedge d x^{j_q}.
\eqno(2.5)$$ where $a_{k, j_1...j_q}(x,y) = a_{k, i_1...i_p,j_1...j_q}(x) 
y^{i_1}...y^{i_p}$. 

The set of differential forms constitutes (similarly as the usual ones) a
{\it Grassmann - Cartan} algebra ${\cal C}= {\cal E} \big( \tilde W
\otimes \Lambda \big)= \bigoplus_{q=0}^{2n} {\cal E}\big( \tilde W \otimes
\Lambda^q\big)$. In this space the multiplication
$\buildrel{\bullet}\over{\wedge}$ is defined by

$$a \buildrel{\bullet}\over{\wedge} b = a_{[j_1\ldots j_p} \bullet b_{l_1 
\ldots l_q]} dx^{j_1} \wedge \ldots \wedge dx^{j_p}\wedge dx^{l_1} 
\wedge \ldots \wedge dx^{l_q}, \eqno(2.6)$$
for all $a = \sum_k \hbar^k a_{k,j_1 \ldots j_p}(x,y) dx^{j_1}\wedge
\ldots \wedge dx^{j_p} \in {\cal E}\big(\tilde{\cal W} \otimes
\Lambda^p\big)$ and $b = \sum_k \hbar^k b_{k, l_1 \ldots l_q}(x,y)
dx^{l_1}\wedge \ldots \wedge dx^{l_q} \in {\cal E}\big ( \tilde{\cal W}
\otimes \Lambda^q \big)$. $a \buildrel{\bullet}\over{\wedge} b$ is defined by the usual wedge
product on ${\cal M}$ and the product $\bullet$ in the Weyl algebra.

A very useful concept is that of central forms. In order to define it
first consider the commutator defined on the sections ${\cal E}(\tilde
{\cal W})$ {\it i.e.}, for all $a \in {\cal E} \big ( \tilde W \otimes
\Lambda^{q_1}\big)$ and $b \in {\cal E} \big( \tilde{\cal W} \otimes
\Lambda^{q_2} \big)$ we define

$$ [ a, b] \equiv a \buildrel{\bullet}\over{\wedge} b - (-1)^{q_1q_2} b \buildrel{\bullet}\over{\wedge} a. \eqno(2.7) $$
Thus a form $a \in {\cal E} \big ( \tilde {\cal W} \otimes \Lambda\big )$
is said to be central, if for any $b \in {\cal E} \big ( \tilde{\cal W}
\otimes \Lambda\big)$ the commutator (2.7) vanishes. The set of central
forms is designed by ${\cal Z} \otimes \Lambda$. Here ${\cal Z}$ coincides
with the algebra of quantum observables\foot{ The algebra of quantum
observables can be defined introducing an associative product operation
$*$ on the vector space ${\cal Z}$ of functions $a(x,\hbar) =
\sum_{k=0}^{\infty} \hbar^k a_k(x)$ and $b(x,\hbar) = \sum_{k=0}^{\infty}
\hbar^k b_k(x)$, with $a_k(x), b_k(x)  \in C^{\infty}({\cal N})$ (${\cal
N}$ is an `internal' manifold). The product $*$ is defined by $a*b=c =
\sum_{k=0}^{\infty}c_k(x)$ for all $a,b,c \in {\cal Z}$ satisfying the
properties (i)-. $c_k$ are polynomials in $a_k$ and $b_k$ and their
derivatives. (ii)-. $c_0(x) = a_0(x) b_0(x)$. (iii)-. $[a,b] \equiv a*b -
b*a = +i \hbar \{a_0,b_0\}_P + ... \ \ \ .$}. In order to be more precise
${\cal Z}$ is a linear space whose elements are

$$ a=a(x, \hbar)= \sum_{k=0}^{\infty} \hbar^k a_k(x), \eqno(2.8)$$
where $a_k(x) \in C^{\infty}({\cal M})$.

Let $a (x, y, \hbar)$ be an element of ${\cal E} (\tilde {\cal W})$, we
define the symbol map $\sigma: {\cal E} (\tilde {\cal W}) \to {\cal Z}, \
\
a (x, y, \hbar) \mapsto a (x,0, \hbar)$, that is the map $\sigma$ is the
projection of ${\cal E}(\tilde{\cal W})$ onto ${\cal Z}$.

\vskip 1truecm
\section{Differential Operators}

One can define some important differential operators. The operator $
\delta:  {\cal E} \big ( \tilde {\cal W}_p \otimes \Lambda^q\big) \to
{\cal E} \big( \tilde {\cal W}_{p-1} \otimes \Lambda^{q+1}\big)$ defined
by

$$ \delta a \equiv d x^k \wedge {\partial a\over \partial y^k}
\eqno(2.9)$$
and its dual operator $\delta^{\bullet}: {\cal E} \big( \tilde {\cal W}_p
\otimes 
\Lambda^q \big) 
\to {\cal E} \big( \tilde {\cal W}_{p+1} \otimes \Lambda^{q-1}\big) $
defined by

$$ \delta^{\bullet}a \equiv y^k {\partial \over \partial x^k} \rfloor a.
\eqno(2.10)$$
for all $a \in {\cal E} \big( \tilde{\cal W}_p \otimes \Lambda^q\big),$
where $\rfloor$ stands for the contraction.

A useful definition is that of the differential operator $\delta^{-1}$
acting on a monomial $y^{i_1} y^{i_2}...y^{i_p} dx^{j_1} \wedge dx^{j_2}
\wedge ... \wedge dx^{j_q}$. $\delta ^{-1}$ is defined as

$$ \delta^{-1}:= {\delta^{\bullet} \over (p+q)}, \ \ \ \ p+q>0$$
and $\delta^{-1} =0,$ for $p+q =0$.

The operators $\delta$ and $\delta^{\bullet}$ satisfy several properties very 
similar to those for the usual differential and co-differential; for 
instance, there exists an analogue of Hodge-de Rham decomposition [24].

\subsection{Symplectic Connection}

Assume the existence of a torsion-free connection on ${\cal M}$ which
preserves its symplectic structure. This connection is known as {\it
symplectic connection} ${\partial_i}$. 

The definition of an operator which do not change the degree of the Weyl
algebra and only changes the degree of differential forms is also
possible. This operator is a connection defined in the bundle $\tilde{\cal
W}$ as $ \partial: {\cal E} \big( \tilde{\cal W} \otimes \Lambda^q\big) 
\to {\cal E} \big( \tilde{\cal W} \otimes \Lambda^{q+1}\big) $ and is
defined in terms of the symplectic connection as follows

$$ \partial a \equiv d x^i \wedge \partial_i a.
\eqno(2.11)$$
In Darboux local coordinates this connection is written 
as 

$$ \partial a = d a + {1 \over i \hbar} [\Gamma , a] \eqno(2.12) $$ 
where $\Gamma = {1\over 2} \Gamma_{ijk}
y^i y^j d x^k$ is a local one-form with values in ${\cal E} \big( \tilde
{\cal W} \big),$ $\Gamma_{ijk}$ are the symplectic connection's
coefficients, $d = d x^i \wedge {\partial \over \partial x^i}$ and 
$\partial_i$ is the covariant derivative  on ${\cal M}$ with respect to
${\partial \over \partial x^i}.$ 

The connection $\partial$  satisfies the following properties:

$$\partial (a \buildrel{\bullet}\over{\wedge}  b) = \partial a \buildrel{\bullet}\over{\wedge}  b + (-1)^{q_1} a \buildrel{\bullet}\over{\wedge} b
\eqno(2.13a)$$

$$\partial (\phi \bullet a)= d\phi \buildrel{\bullet}\over{\wedge} b + (-1)^q \phi \bullet \partial a
\eqno(2.13b)$$ for all $\phi \in {\cal E}(\Lambda^q)$ and $ a \in {\cal E}
(\tilde {\cal W} \otimes \Lambda^{q_1}), b \in {\cal E} \big (\tilde {\cal
W} \otimes \Lambda^{q_2}\big)$.

Following Fedosov, we define a more general connection $D$ in the Weyl
bundle $\tilde{\cal W}$ as follows

$$ Da = \partial + {1 \over i \hbar} [\gamma,a], \eqno(2.14) $$
where $\gamma \in {\cal E}(\tilde{\cal W} \otimes \Lambda^1)$ is globally
defined on ${\cal M}$. The {\it curvature} of the connection $D$ is given
by 

$$ {1\over i \hbar} \Omega = {1 \over i \hbar}(R + \partial \gamma + {1
\over
i \hbar} \gamma^2), \eqno(2.15) $$ with the normalizing condition
$\gamma_0
= 0$. Here $R$ is defined by $R:= {1 \over 4} R_{ijkl}y^iy^j dx^k \wedge
dx^l$ where $R_{ijkl}$ is the curvature tensor of the symplectic
connection. In [24] it was shown that for any section $a \in {\cal
E}(\tilde{\cal W}\otimes \Lambda)$ we have

$$D^2a =  {1 \over i \hbar} [\Omega, a]. \eqno(2.16)$$

\subsection{Abelian Connection}

One very important definition is that of the {\it Abelian connection}. A
connection $D$ is Abelian if for any section $a \in {\cal E}(\tilde{\cal
W} \otimes \Lambda)$

$$ D^2 a =  {1 \over i \hbar} [ \Omega, a] = 0. \eqno(2.17)$$
From Eq. (2.7) one immediately sees that the curvature of the Abelian
connection, $\Omega$, is central.

In Fedosov's paper the Abelian connection takes the form

$$D = -\delta + \partial + {1 \over i \hbar} [r, \cdot], \eqno(2.18)$$
where $\partial$ is a fixed symplectic connection and $r \in {\cal
E}(\tilde{\cal W}_3 \otimes \Lambda^1)$ a globally defined one-form with
the Weyl normalizing condition $r_0 = 0$. This connection has curvature

$$\Omega = - {1 \over 2} \omega_{ij} dx^i \wedge dx^j + R - \delta r +
\partial r + { 1\over i \hbar} r^2 \eqno(2.19)$$
with 
$$\delta r = R + \partial r + {1 \over i \hbar} r^2. \eqno(2.20)$$
This last equation has  a {\it unique} solution satisfying the condition 
$$ \delta^{-1} r = 0. \eqno(2.21)$$

By iterative techniques one can finally construct $r$ and therefore the
Abelian connection $D$. Thus we have

$$r = {1 \over 8} R_{ijkl}y^iy^jy^kdx^l + {1\over 20} \partial_m
R_{ijkl}y^iy^jy^ky^mdx^l + ... \ , \eqno(2.22)$$ 
where $\partial_m$ is a covariant derivative withy respect to ${\partial
\over \partial x^m}.$

\subsection{Algebra of Quantum Observables}

Now consider the subalgebra ${\cal E}(\tilde{\cal W}_D)$ of ${\cal
E}(\tilde{\cal W})$ consisting of {\it flat sections} {\it i.e.}

$$ {\cal E}(\tilde{\cal W}_D) = \{a\in {\cal E}(\tilde{\cal W}) | Da =
0\}. \eqno(2.23)$$
This subalgebra is called {\it the algebra of Quantum Observables}.

Now an important theorem is:
\vskip .5truecm
\noindent {\bf Theorem}(Fedosov [24]) For any $a_0 \in {\cal Z}$ there
exists a unique section $a\in {\cal E}(\tilde{\cal W}_D)$ such that
$\sigma(a) = a_0$. 

As a direct consequence of this theorem we can construct a section $a \in
{\cal E}(\tilde{\cal W}_D)$ by its symbol $a_0 = \sigma(a)$

$$a = a_0 + \partial a_0 y^i + {1 \over 2} \partial_i \partial_j a_0 y^i
y^j + {1 \over 6} \partial_i \partial_j\partial_k a_0 y^i y^j y^k - {1
\over 24} R_{ijkl} \omega^{lm} \partial_m a_0 y^i y^j y^k + ... \ .
\eqno(2.24)$$ 
In the case when the phase space is flat $R_{ijkl} = 0$ the last equation
reads

$$ a = \sum_{k=0}^{\infty} {1 \over k!} ( \partial_{i_1} \partial_{i_2}
... \partial_{i_k} a_0) y^{i_1} y^{i_2} ...y^{i_k}. \eqno(2.25)$$

The last theorem states that there exists the bijective map

$$\sigma: {\cal E}(\tilde{\cal W}_D) \to {\cal Z}. \eqno(2.26) $$
Therefore there exists the inverse map $\sigma^{-1}: {\cal Z} \to {\cal
E}(\tilde{\cal W}_D)$ Is possible to use this bijective map to recover the
Moyal product $*$ in ${\cal Z}$.

$$ a_0 * b_0 = \sigma (\sigma^{-1}(a_0) \bullet \sigma^{-1}(b_0)).
\eqno(2.27) $$

\vskip 1truecm

\section {A definition of Trace on the Weyl algebra on ${\bf R}^{2n}$}

In order to work with a variational principle which involves Moyal geometry
we would like to get a definition of trace. In the case ${\cal M} =
{\bf R}^{2n}$ with the standard symplectic structure

$$ \omega = \sum^{2n}_{j=1} d p_j \wedge d q_j. \eqno(2.28)$$
Here the Abelian connection $D$ in $\tilde{\cal W}\buildrel{\pi}\over{\to}
{\bf R}^{2n}$ is $D = - \delta + d$.

In this case the product $\star$ coincides with the usual Moyal product
[8],

$$ \sigma (a \bullet b) ={\rm exp} \bigg( {+i\hbar \over 2} \omega^{ij}
{\partial\over \partial x^i} {\partial \over \partial y^j} \bigg) \sigma(a
(x, \hbar))  \sigma( b (z, \hbar)) |_{z=x} $$

$$ = \sum^\infty_{k=0} \bigg(+{i\hbar\over 2} \bigg)^k {1 \over k!}
\omega^{i_1 j_!} \ldots \omega^{i_k j_k} \ {\partial^k a_0\over \partial
x^{i_1}\ldots \partial x^{j_1}, } \ \ {\partial^k b_0 \over \partial
x^{j_1}\ldots \partial x^{j_k}} \eqno(2.29)$$ 
where $a_0:= \sigma(a)$ and $b_0 = \sigma(b)$.

The trace in the Weyl algebra ${\cal E}(\tilde{\cal W}_D)$ over $ {\bf
R}^{2n}$ is the linear functional on the ideal ${\cal E}({\cal W}_D^{\rm
Comp})$ over ${\cal M} = {\bf R}^{2n}$
(which consists of the flat sections with compact support) given by

$$ tr \ a =  \int_{{\bf R}^{2n}} \sigma (a) 
{\omega^n \over n!} \eqno(2.30)$$
where $\sigma(a)$ means the projection on the center
$\sigma(a(x,y,\hbar)) := a(x,0,\hbar).$

This definition of the trace satisfies a series of useful properties

$$ a) \ \ \ tr (a \bullet b) = tr (b \bullet a) \eqno(2.31a) $$

$$ b) \ \ \ tr (b) = tr (A_f b)  \eqno(2.31b)$$ 

for all the sections $a \in {\cal E}(\tilde{\cal W}_D)$, $b \in {\cal
E}(\tilde{\cal W}_D^{comp}).$ In last equation, $A_f$ is an isomorphism
$A_f: {\cal E}(\tilde{\cal W}_D^{comp})({\cal O}) \to {\cal
E}(\tilde{\cal W}_D^{comp})(f({\cal O})),$ where $f$ is a symplectic
diffeomorphism.

On the other hand it is possible to construct a trace on the sections
algebra ${\cal E}(\tilde{\cal W}_D)$ of the Weyl bundle over an arbitrary
symplectic manifolds ${\cal M}$. Although this definition satisfies the
property (2.31), unfortunately it is too formal and we don't consider it
here.

\vskip 2truecm


\chapter{\bf Moyal Deformation of Self-dual Gravity}

We now recall some results of Refs. [20,21]. We first review the chiral
model approach to SDG and its Moyal deformation [21]. Then we consider the
Yang-Mills approach to SDG [20]. 

\section{ The Principal Chiral Model Approach to Self-dual Gravity}

We start with the principal chiral model approach to SDG \`a la Husain
[9], Park [7] and Ward [6].  Husain has shown the equivalence between SDG
and sdiff$(\Sigma)$-valued principal chiral model. Since then some
solutions of this model has been found in terms of harmonic maps [28].

In Ref. [21] we found that the Moyal deformation of Park-Husain (P-H) 
heavenly equation can be obtained from the operator algebra valued
two-dimensional principal chiral model. To this end the WWM-formalism has
been employed. We have finally reproduced the P-H heavenly equation by
taking the limit $\hbar \to 0$, instead of $N \to \infty$.

\subsection{The Principal Chiral Model}

The ${\cal G}$-valued principal chiral equations on a two-dimensional
simply connected manifold $\Omega$ with local Cartesian coordinates
$\{x,y\}$ read

$$\partial_x A_y - \partial_y A_x + [A_x,A_y] = 0, \eqno(3.1a)$$ 

$$\partial_x A_x + \partial_y A_y = 0, \eqno(3.1b)$$ 
where $A_{\mu} \in {\cal G} \otimes C^{\infty}({\Omega})$, $\mu \in
\{x,y\}$, stand for the chiral potentials and ${\cal G}$ is a Lie algebra
of the Lie group {\bf G}. 

One can proceed as follows. From $(3.1a)$ it follows that $A_{\mu}$,
${\mu} \in \{x,y\},$ is of the pure gauge form, {\it i.e.}, there exists a
${\bf G}$-valued function $g = g (x,y)$ such that

$$ A_{\mu} = g^{-1} \partial_{\mu} g. \eqno(3.2)$$ 
Substituting $(3.2)$ into $(3.1b)$ we get the principal chiral equations
$$ \partial_{\mu} \big( g^{-1} \partial_{\mu} g \big) = 0. \eqno(3.3)$$
(Summation over $\mu$ is assumed.) 

Chiral equations are the dynamical equations for the fields $g : \Omega
\to {\bf G}$ which under specific boundary conditions are called {\it
harmonic maps} [22].

It is very easy to see that the Lagrangian for equations of motion (3.3) 
reads (we asumme that ${\cal G}$ is semisimple) 

$${\cal L}_{Ch} = - c {\rm Tr} \big \{ (g^{-1} \partial_{\mu}
g)(g^{-1}\partial_{\mu} g) \big \}$$ $$ = c {\rm Tr} \big \{
(\partial_{\mu} g)(\partial_{\mu}g^{-1}) \big \}, \eqno(3.4)$$
where $c >0$ is a constant and `Tr' is an invariant form on the Lie 
algebra ${\cal G}.$

Let $ \hat {\bf G}$ be some Lie group of linear operators acting on the
Hilbert space $L^2(\Re^1)$ and let $\hat{\cal G}$ be the Lie algebra of $
\hat {\bf G}$. Consider the $ \hat {\bf G}$ principal chiral model.  The
principal chiral equations read now

$$\partial_x \hat{A}_y - \partial_y \hat{A}_x + [\hat{A}_x,\hat{A}_y] = 0,
\eqno(3.5a)$$ 

$$ \partial_x \hat{A}_x + \partial_y \hat{A}_y = 0,
\eqno(3.5b)$$ 
where $\hat{A}_{\mu} = \hat{A}_{\mu}(x,y) \in \hat{\cal G}
\otimes C^{\infty} ({\Omega}), \ \mu \in \{x,y\}.$

From the constraint $(3.5a)$ one infers that $$ \hat{A}_{\mu} =
\hat{g}^{-1} \partial_{\mu} \hat{g},\eqno(3.6)$$ where $\hat{g} =
\hat{g}(x,y)$ is some $\hat{\bf G}$-valued function on $\Omega$.
Substituting (3.6) into $(3.5b)$ we get the principal chiral equations

$$ \partial_{\mu} \big( \hat{g}^{-1} \partial_{\mu} \hat{g} \big) = 0.
\eqno(3.7)$$

Within WWM-formalism we can transform the above equation into a new
equation defined on the four manifold ${\cal K}^4 = \Omega \times \Sigma$,
being $\Sigma \subset {\bf R}^2$. 

The Weyl {\it correspondence} ${\cal W}^{-1}$ leads  from $\hat{g} = 
\hat{g}(x,y)$ to the function on ${\cal K}^4$, $g = g(x,y,p,q;\hbar)$, 
{\it i.e.}, $ g = {\cal W}^{-1}(\hat{g})$, according to the formula
(compare with [17-21,23])

$$ g = g(x,y,p,q) := {\cal W}^{-1}\big( \hat{g}(x,y)\big) = \int_{- 
\infty}^{ + \infty} \big< q - {\xi \over 2} \big | \hat{g} \big | q  + 
{\xi \over 2} \big > {\rm exp} \big(  {i p \xi \over \hbar} \big) d \xi 
\eqno(3.8) $$  
from Eq. (3.7) one can infer that the above function fulfills the
following equation 

$$ \partial_{\mu} \bigg( g^{- \buildrel{*}\over{1}} * \partial_{\mu} g
\bigg) = 0, \eqno(3.9)$$
where $*$ stands for the {\it Moyal} $*${\it -product} (see (3.24)) and 
$g^{- \buildrel{*}\over{1}}$ denotes the
inverse of $g$ in the sense of the Moyal $*$-product {\it i.e.}, 

$$ g^{- \buildrel{*} \over{1}} * g = g * g^{- \buildrel{*}\over{1}} = 1. 
\eqno(3.10)$$

Comparing (3.9) with (3.7) we can say that the function $ g = {\cal
W}^{-1}\big( \hat{g}(x,y)\big)$ defines a {\it harmonic map} $ g: \Omega
\to {\bf G}_*$, being ${\bf G}_* := {\cal W}^{-1}(\hat {\bf G})$\foot{ In
a sense the Moyal bracket algebra can be considered to be an infinite
dimensional matrix Lie algebra. Especially interesting is the case when
the group ${\bf G}_*$ is a subgroup of ${\bf U}_*$, where $ {\bf U}_* :=
\{ f = f(p,q) \in C^{\infty}({\bf R}^2); \ f* \bar{f} = \bar{f} * f = 1\}$; 
(the bar stands for the complex conjugation.). It means that $\hat{\bf G}
= {\cal W}({\bf G}_*)$ is a subgroup of the group $\hat{\bf U}$ of unitary
operators acting on $L^2({\bf R}^1).$ Now one quickly finds that if $g =
g(x,y,p,q)$ is an ${\bf U}_*$-valued function, then $ g^{-
\buildrel{*}\over{1}} * \partial_{\mu} g = \bar{g} * \partial_{\mu} g$ is
{\it pure imaginary}.}.

The Lagrangian associated to Eq. (3.9) can be written as [21]

$${\cal L}^{\prime (M)}_{SG} := - {{\hbar}^2 \over 2} \big(g^{-
\buildrel{*} \over{1}} * \partial_{\mu} g \big)* \big( g^{- \buildrel {*}
\over{1}} * \partial_{\mu} g \big), \eqno(3.11)$$
or equivalently by the Lagrangian 

$$ {\cal L}^{\prime \prime (M)}_{SG} = {{\hbar}^2 \over 2}
\big(\partial_{\mu} g \big)*\big(\partial_{\mu} g^{-\buildrel{*} \over{1}}
\big). \eqno(3.12)$$

\subsection{The Moyal Deformation of Park-Husain Heavenly Equation}

To proceed further, we need the Moyal deformation of P-H 
version of SDG.

We start with Eq. $(3.1b)$. This equation says that there exists a scalar
function $\theta = \theta (x,y) \in {\cal G} \otimes C^{\infty}(\Omega)$
such that
 
$$ A_x = - \partial_y \theta, \ \ \ \ \ {\rm and} \ \ \ \ \ A_y =
\partial_x \theta. \eqno(3.13)$$ 

Inserting Eqs. (3.13) into $(3.1a)$ one gets the principal chiral 
equations to read

$$ \partial_x^2 \theta + \partial_y^2 \theta + [\partial_x \theta,
\partial_y \theta] = 0. \eqno(3.14)$$

Under the assumption that the algebra ${\cal G}$ is semisimple one can 
construct a Lagrangian leading to $(3.14)$ as follows

$$ {\cal L}_{Ch} := c' {\rm Tr} \bigg \{ {1\over 3} \theta [\partial_x
\theta, \partial_y \theta] - {1 \over 2} \bigg( (\partial_x \theta)^2 +
(\partial_y \theta)^2 \bigg) \bigg \}, \eqno(3.15)$$ 
where $c' > 0$ is a constant and `Tr' is defined as (3.4).

Similarly from $(3.5b)$ it follows that there exists the operator$-$valued
scalar function $\hat{\theta} = \hat {\theta}(x,y) \in \hat {\cal G}
\otimes C^{\infty} (\Omega)$ such that

$$ \hat{A}_x = - \partial_y \hat{\theta}, \ \ \ \ \ {\rm and} \ \ \ \ \  
\hat{A}_y = \partial_x \hat{\theta}. \eqno(3.16)$$

The analog of Eq. (3.14) is $$ \partial^2_x \hat \theta + \partial^2_y
\hat \theta + [\partial_x \hat \theta, \partial_y \hat \theta] = 0.
\eqno(3.17)$$

Now, it is convenient to define a new operator-valued function
$\hat{\Theta} = \hat{\Theta}(x,y) \in \hat{\cal G} \otimes
C^{\infty}(\Omega)$ by

$$ \hat{\Theta} := i \hbar \hat{\theta}. \eqno(3.18)$$

Thus, by (3.17),  $\hat{\Theta}$ satisfies the following equation

$$ \partial^2_x \hat \Theta + \partial^2_y \hat \Theta + {1 \over
i{\hbar}} [\partial_x \hat \Theta, \partial_y \hat \Theta] = 0.
\eqno(3.19)$$
Then we put 

$$ \Theta = \Theta(x,y,p,q,\hbar) := {\cal W}^{-1}\big( 
\hat{\Theta}(x,y)\big) = \int_{- \infty}^{ + \infty} \big< q - {\xi 
\over 2} \big | \hat{\Theta} \big | q  + {\xi \over 2} \big > {\rm exp} 
\big(  {i p \xi \over \hbar} \big) d\xi. \eqno(3.20) $$
It is clear that $\Theta$ satisfies the {\it Moyal deformation of the 
P-H heavenly equation}

$$ \partial^2_x \Theta + \partial^2_y \Theta + \{\partial_x \Theta,
\partial_y \Theta \}_M = 0, \eqno(3.21)$$ 
where the bracket $\{\cdot,\cdot \}_M$ denotes the {\it Moyal bracket}
{\it i.e.},

$$ \{f_1,f_2\}_M := {1 \over i {\hbar}} (f_1 * f_2 - f_2* f_1) = f_1 {2
\over {\hbar}} {\rm sin}({{\hbar}\over 2} \buildrel {\leftrightarrow}
\over {\cal P}) f_2, \eqno(3.22)$$

$$\buildrel {\leftrightarrow}\over {\cal P} := {\buildrel
{\leftarrow}\over{\partial} \over \partial q} 
{\buildrel{\rightarrow}\over { \partial} \over
\partial p} - {\buildrel{\leftarrow}\over{\partial} \over \partial p}
{\buildrel{\rightarrow}\over {\partial}  \over \partial q}; \ \ \ 
f_1=f_1(x,y,p,q), \ \ \ f_2 = f_2(x,y,p,q). \eqno(3.23)$$
The {\it Moyal $*$-product} is defined by

$$ f_1*f_2 := f_1 {\rm exp} ({i {\hbar} \over 2}
\buildrel{\leftrightarrow}\over{\cal P})f_2. \eqno(3.24)$$ 
If the functions $f_1$ and $f_2$ are independent of $\hbar$, then

$$\lim_{{\hbar}\to 0} f_1 * f_2 = f_1f_2, \ \ \ \ \ \lim_{{\hbar} \to 0}
\{f_1,f_2\}_M = \{f_1,f_2\}_P := f_1 \buildrel {\leftrightarrow}\over
{\cal P} f_2, \eqno(3.25)$$ 
where $\{\cdot,\cdot \}_P$ denotes the Poisson bracket.

As it has been shown in [21] the Lagrangian leading to Eq. (3.21) reads 

$$ {\cal L}^{(M)}_{SG} = - {1 \over 3} \Theta * \{\partial_x \Theta,
\partial_y \Theta \}_M + {1 \over 2} \bigg( (\partial_x
\Theta)*(\partial_x \Theta) + (\partial_y \Theta)*(\partial_y \Theta)
\bigg). \eqno(3.26)$$ 

\vskip 1truecm

\subsection{The Park-Husain Heavenly Equation}

Assume now that the function $\Theta$ is analytic in $\hbar$, {\it i.e.,}
[14]

$$ \Theta = \sum_{n = 0}^{\infty} {\hbar}^n \Theta_n, \eqno(3.27)$$ 
where $\Theta_n = \Theta_n(x,y,p,q),$ $n=0,1,...,$ are independent of
$\hbar$. If $\Theta$ is a solution of (3.21), then by (3.27) one concludes
that the function $\Theta_0$ satisfies {\it P-H heavenly equation}
[21]

$$ \partial^2_x \Theta_0 + \partial^2_y \Theta_0 + \{ \partial_x \Theta_0,
\partial_y \Theta_0 \}_P = 0. \eqno(3.28)$$ Moreover, the Lagrangian
${\cal L}_{SG}$ leading to Eq. (3.28) can be quickly found to read

$${\cal L}_{SG} = \lim_{\hbar \to 0} {\cal L}^{(M)}_{SG} = - {1 \over 3}
\Theta_0 \{ \partial_x \Theta_0, \partial_y \Theta_0 \}_P + {1\over 2}
\bigg( (\partial_x \Theta_0)^2 + (\partial_y \Theta_0)^2 \bigg)
\eqno(3.29)$$
(compare with the Lagrangian well known in SDG [29]).

Therefore, {\it self-dual gravity appears to be the $\hbar \to 0$ limit of
the principal chiral model for the Moyal bracket algebra,} or
equivalently, one can interpret self-dual gravity to be the {\it principal
chiral model for the Poisson bracket algebra} [6,7,9]. 

If one is interested in searching for solutions
of P-H heavenly equation (3.28), one must take an explicit  Lie algebra 
homomorphism $\Psi : {\cal G} \to \hat{\cal G}$ 

$$ \hat{\Theta} = \hat{\Theta}(x,y) = i \hbar \theta_a(x,y) \hat{X}_a, 
\eqno(3.30) $$ 
where $ \hat{X}_a := \Psi(\tau_a) $ satisfies Eq. (3.19).
Therefore, the function $\Theta$ defined by (3.20)

$$ \Theta = \Theta(x,y,p,q) = i \hbar \theta_a(x,y) X_a(p,q) $$ 

$$X_a(p,q):= {\cal W}^{-1}(\hat{X}_a) \eqno(3.31)$$ 
fulfills the  Moyal deformation of P-H heavenly equation (3.21). 

Consequently, if $\Theta$ is of the form (3.31) then $\Theta_0$ satisfies
{\it P-H heavenly equation}, (3.28). Moreover, if the Lie group
$\hat{\bf G}$ defined by the Lie algebra $\hat{\cal G}$ appears to be a
subgroup of the group $\hat{\bf U}$ of unitary operators in $L^2({\bf R}^1)$
then the functions $\Theta$ and $\Theta_0$ are real. 

The procedure described here, which leads to the construction of the
solutions to P-H heavenly equation is somewhat speculative. The main
problem is to find representations for interesting Lie algebras and show
how it works in practice. (For su$(2)$ see [21]).

\vskip 1truecm

\section{SDYM Theory Approach to Self-dual Gravity}

We will now describe the su$(N)$ SDYM equations in the flat 4-dimensional
real, simply connected flat manifold $X \subset{\bf R}^4$ with local
coordinates $(x,y,\tilde{x},\tilde{y})$ chosen in such a way that the metric takes the form $dS^2 = 2(dx \otimes_s d\tilde{x} + dy \otimes_s d \tilde{y}).$

Then the su$(N)$ SDYM equations read

$$
F_{xy}  = 0, \ \ \ \ \   F_{\tilde x \tilde y} = 0, \ \ \ \ \ F_{x \tilde 
x} \ + \ F_{y \tilde y} = 0, \eqno(3.32)$$
where, as usually, $F_{\mu \nu} \in  {\rm su}(N) \otimes C^{\infty}
(X), \mu , \nu  \in  \{x, y, \tilde x, \tilde y\}$, stands for the 
Yang-Mills field tensor. 

 In terms of the Yang-Mills potentials $A_\mu \in {\rm su}(N) \otimes 
C^{\infty} (X)$  the  SDYM equations can be 
rewritten in the gauge $A_x = A_y = 0$ as 
$$
\partial_{\tilde x} A_{\tilde y} + \partial_{\tilde y} A_{\tilde x} + 
[A_{\tilde x}, A_{\tilde y} ] = 0, \eqno(3.33a)
$$
$$
\partial_x A_{\tilde x} + \partial_y A_{\tilde y} = 0. \eqno(3.33b)
$$

Now assume that the potentials $A_{\mu}$ are now the anti-hermitian 
operator-valued functions on $ X \subset {\bf R}^4$ acting in a Hilbert
space ${\cal H} = L^2 ({\bf R}^1)$. In this case Eqs. (3.32) give

$$
\partial_{\tilde x} \hat A_{\tilde y} - \partial_{\tilde y} \hat A_{\tilde x} 
+ [\hat A_{\tilde x}, \hat A_{\tilde y} ] = 0, \eqno(3.34a)
$$
$$ 
\partial_x \hat A_{\tilde x} + \partial_y \hat A_{\tilde y} = 0, \eqno(3.34b)
$$
where
$$ 
\hat A^{\dag}_{\tilde x} = - \hat A_{\tilde x},\ \ \ \ \ \ \   \hat 
A^{\dag}_{\tilde y} = - \hat A_{\tilde y}. $$

\vskip 1truecm

\subsection{Moyal Deformation of the Six Dimensional Version of the Second 
Heavenly Equation}

Eq. $(3.34b)$ implies that

$$ \hat{A}_x = - \partial_y \hat{\theta}, \ \ \ \ \ \ \hat{A}_y = 
\partial_x \hat{\theta}, \eqno(3.35) $$
under the condition

$$ \hat{\theta} = \hat{\theta}(x,y,\tilde{x}, \tilde{y}) = - 
\hat{\theta}^{\dag}. $$
It is easy to see from Eq. $(3.34a)$ and (3.35) that

$$ \partial_x \partial_{\tilde x} \hat{\Theta} + \partial_y
\partial_{\tilde y} \hat{\Theta} + {1 \over i\hbar} [\partial_x
\hat{\Theta}, \partial_y \hat{\Theta}] = 0. \eqno(3.36) $$
where $ \hat{\Theta} := i \hbar \theta.$

The above equation can be derived as equation of motion from the 
Lagrangian [20]
 
$$ {\cal L}^{(q)} =  Tr \big \{ 2 \pi  \hbar  \big( - {1 \over 3i 
\hbar} \hat{\Theta} [\partial_x \hat{\Theta}, \partial_y \hat{\Theta}] + 
{1 \over2} [ (\partial_x \hat{\Theta})(\partial_{\tilde x} \hat{\Theta}) 
+  (\partial_y \hat{\Theta})(\partial_{\tilde y} \hat{\Theta})] \big) 
\big\}. \eqno(3.37)$$ 
where `$Tr$' is the standard trace of a linear operator in the 
orthonormal  base \   $\{|\psi>_j \}_{j \in {\bf N}}$ of the Hilbert 
space ${\cal H} = L^2({\bf R}^1)$. 

Using the WWM-formalism it can be shown [20] that the above Lagrangian can
be transformed into a new Lagrangian defined on the six-dimensional
manifold $ {\cal K}^6 = X \times \Sigma$, $\Sigma = {\bf R}^2$. Let
$\{p,q\}$ be the local coordinates on $\Sigma$.  This new Lagrangian reads

$${\cal L}^{(M)} = - {1 \over 3}\Theta * \{\partial_x \Theta, \partial_y 
\Theta \}_M + {1 \over 2}[(\partial_x \Theta)*(\partial_{\tilde x} 
\Theta) + (\partial_y \Theta)*(\partial_{\tilde y} \Theta)]. \eqno(3.38)$$

Furthermore the {\it Weyl correspondence} ${\cal W}^{-1}$ leads from
$\hat{\Theta} = \hat {\Theta}(x,y, \tilde{x},\tilde{y})$ to the function
$\Theta = \Theta(x,y,\tilde{x},\tilde{y},p,q, \hbar)$, $ \Theta = {\cal
W}^{-1}(\hat{\Theta})$, defined on $ X \times \Sigma$ (here $\Sigma =
{\bf R}^2$) according to the formula

$$ \Theta = \Theta(x,y,\tilde{x},\tilde{y},p,q, \hbar) := \int_{- \infty}^{+ 
\infty} {\rm exp} 
\big( {ip \xi \over {\hbar}} \big) < q- {\xi\over 2}| \hat {\theta}| q +
{\xi \over 2}> d\xi. \eqno(3.39)$$

In Ref. [20] it has been also shown that $\Theta \in C^{\infty}( X \times {\bf R}^2)$ satisfies the {\it Moyal deformation of the
six-dimensional version of the heavenly equation}

$$ \partial_x \partial_{\tilde x}  \Theta + \partial_y \partial_{\tilde 
y}  \Theta + \{ \partial_x \Theta,
\partial_y \Theta \}_M = 0, \eqno(3.40)$$ 
which is, of course, the equation of motion of the Lagrangian (3.38).

Taking the limit $\hbar \to 0$ in the Lagrangian (3.38)  one obtains

$$ {\cal L}_{\infty} = - {1\over 3} \Theta 
\{\partial_x \Theta,\partial_y \Theta \}_P + {1\over 2} [(\partial_x 
\Theta)(\partial_{\tilde x} \Theta) + (\partial_y 
\Theta)(\partial_{\tilde y} \Theta)], \eqno(3.41) $$
which has as Euler-Lagrange equation
precisely the {\it six-dimensional version of the second heavenly 
equation}\foot{ The six-dimensional version of the second heavenly 
equation can be obtained from the su$(N)$-SDYM equations  
by taking the large-$N$ limit, $N\to \infty.$}

$$\partial_x \partial_{\tilde x} 
\Theta  +  \partial_y \partial_{\tilde 
y} \Theta + \{\Theta_x,\Theta_y\}_P = 0, \eqno(3.42) $$
where $\Theta = \Theta(x,y,\tilde{x},\tilde{y},p,q)$ is a smooth function on 
the six-dimensional manifold ${\cal K}^6.$ As it has been demonstrated in  
Ref. [20], the different (but equivalent) versions of the heavenly 
equation (for instance, first and second heavenly equations, Grant's 
equation [30], P-H equation  and the evolution form of the second 
heavenly equation [31]) are symmetry reductions of Eq. (3.42).

\vskip 2truecm


\chapter{\bf A Geometry  Associated  with the Moyal Deformation of 
Self-dual Gravity}

The purpose of this section is to reformulate some results of   
[20,21,23], in terms of non-commutative geometry developed by Fedosov [24] and described in Sec. 2. 

\section{Geometry of Deformation Quantization Associated with the 
Principal Chiral Model Approach to Self-dual Gravity}

\subsection{Equations}

Let us now work out the principal chiral model (described in Section 
(3.1)) in geometrical terms. First of all note that Eqs. $(3.1ab)$ can be 
normally written as

$$ F = dA + A\wedge A = 0, \eqno(4.1a)$$

$$ d \star A = 0 \eqno(4.1b)$$
where $\star$ is the standard Hogde operator and $A \in {\cal E} \big({\cal
G}\otimes 
\Lambda^1\big)$ is the connection one form. 

The corresponding Eqs. (3.2) and (3.3) are

$$ A = g^{-1} dg, \eqno(4.2a)$$
$$ d \star (g^{-1}dg) = 0. \eqno(4.2b)$$
The first equation $(4.1a)$ is the condition of {\it flat connection} 
and the second one is the equation of motion.

In coordinates $(x,y) \in \Omega$ 

$$ A = A_x dx + A_y dy, \eqno(4.3)$$
with 

$$ A_{\mu}(x,y)= \sum_{a=0}^{dim \ G} A^a_{\mu}(x,y) \tau_a \in {\cal G} \otimes C^{\infty}(\Omega), \ \ \mu=x,y. \eqno(4.4)$$

Now we generalize this gauge connection from ${\cal G}$-valued connection one-form (4.3) to the corresponding ${\cal E}(\tilde{\cal W}_D)$-valued connection one-form 

$$ \tilde{A} = \tilde{A}_x dx + \tilde{A}_y dy, \eqno(4.5)$$
with 

$$ \tilde{A}_{\mu}= \tilde{A}_{\mu}(x,y,p,q;\hbar)= a_{\mu} + \partial_i a_{\mu} y^i + {1\over 6} \partial_i \partial_j \partial_k a_{\mu} y^i y^j y^k - {1\over 24} R_{ijkl}\omega^{lm} \partial_m a_{\mu} y^i y^j y^k + \cdots, \eqno(4.6)$$
for the case of non-flat phase-space (with $y^1\equiv p$, $y^2 \equiv q)$. While that for the flat case we have

$$\tilde{A}_{\mu}(z,\bar{z}) = \sum_{k =0}^{\infty} {1\over k!} \big(\partial_{i_1} \partial_{i_2} \cdots \partial_{i_k} a_{\mu} \big) y^{i_1} y^{i_2} \cdots y^{i_k}. \eqno(4.7)$$
In the above formulas $ a_{\mu} = a_{\mu}(x,y,p,q;\hbar).$

The mentioned correspondence also implies that Eqs. $(4.1ab)$ have a 
counterpart in terms of Fedosov's geometry

$$ \tilde{F} = d\tilde{A} + \tilde{A} \buildrel{\bullet}\over {\wedge} 
\tilde{A} = 0, \eqno(4.8a)$$

$$ d \star \tilde{A} = 0,  \eqno(4.8b) $$
where $\tilde{A} \in {\cal E}({\cal E}(\tilde{\cal W}_D) \otimes \Lambda ^1)$ and 
$\tilde{F} \in {\cal E}({\cal E}(\tilde{\cal W}_D)\otimes \Lambda^2).$
Equations $(4.2ab)$ are written now as 

$$\tilde{A} = g^{- \buildrel{\bullet}\over{1}} \bullet dg, \eqno(4.9a)$$

$$ d\star(g^{- \buildrel{\bullet}\over{1}} \bullet dg), \eqno(4.9b)$$

\vskip 1truecm

\subsection{Lagrangian}

Now we will show that  Eqs. $(4.8ab)$ can be obtained from a variational 
principle from a Lagrangian of the standard principal chiral model.

First we recall that the action which gives Eqs. ($4.1ab$) reads

$$ S = \int_{\Omega} {\cal L} $$
where
$$ {\cal L} ={1 \over 2}  {\rm Tr} ( g^{-1} dg \wedge  \star g^{-1}dg), \eqno(4.10)$$
where $g: \Omega \to {\bf G}$ and $d$ is the exterior differential on 
$\Omega$ 
{\it i.e.} $d = dx \partial_x + dy \partial_y$ 
and Tr is an invariant form on the Lie algebra of 
{\bf G}, $Lie({\bf G}) = {\cal G}$. Here we have assumed that {\bf G} is 
semisimple. 

The above action can be generalized  to Fedosov's geometry as follows

$$ S^{\bullet} = \int_{\Omega} {\cal L}^{\bullet}$$
where

$$ {\cal L}^{\bullet}= - { \hbar^2 \over 2}tr\big( \tilde{A} \buildrel{\bullet}\over{\wedge} \star \tilde{A} \big)$$

$$ = - {\hbar^2 \over 2} tr \big(g^{- \buildrel{\bullet}\over{1}} \bullet dg 
\buildrel{\bullet}\over{\wedge} \star g^{- \buildrel{\bullet}\over{1}}\bullet dg \big). \eqno(4.11) $$
where `$tr$' is the Fedosov's trace and $g: \Omega \to {\bf G}_{\bullet}$ is the generalized gravitational uniton (see [23]).

In the case of flat phase-space $R_{ijkl} = 0$, the trace can be expressed by (2.30)

$$ {\cal L}^{\bullet} = - {\hbar^2 \over 2} tr(\tilde{A}\buildrel{\bullet}\over{\wedge} \tilde{A})= - {\hbar^2 \over 2} \int_{{\bf R}^2} \sigma \big(\tilde{A} \buildrel{\bullet}\over{\wedge} \star \tilde{A} \big) dp \wedge dq,  $$

$$ = - {\hbar^2 \over 2} \int_{{\bf R}^2} \sigma\big(g^{- \buildrel{\bullet}\over{1}} \bullet dg 
\buildrel{\bullet}\over{\wedge} \star g^{- \buildrel{\bullet}\over{1}}\bullet dg \big) dp \wedge dq. \eqno(4.12) $$
This corresponds to (3.11).

\vskip 1truecm

\section{ Moyal Deformation of SDYM Theory}

\subsection{ Moyal Deformation of Yang and Donaldson-Nair-Schiff Equations}

First of all, we will apply our method described in [20,21,23] to the 
well known Yang and Donaldson-Nair-Schiff equations. We will find that 
these equations admit Moyal deformations via Fedosov's geometry.

Some years ago C.N. Yang found that su$(2)$-SDYM equations (in a particular 
gauge), always are related to a principal chiral model on a 
four-dimensional flat submanifold $X$ of ${\bf R}^4$ [30] (see 
also [20]). Thus, SDYM equations 
$(3.2ab)$ can be written in apprppriate real coordinates $\{x,y, \tilde{x}, \tilde{y} \}$ as follows

$$ \partial_x\big(g^{-1} \partial_{\tilde{x}} g \big) + \partial_y \big(g^{-1} \partial_{\tilde{y}} g \big) = 0. \eqno(4.14) $$
Equation (4.14) is called {\it Yang equation}.

Proceeding in a similar way as in the above section, we find the equation

$$ \partial_x \big( g^{- \buildrel{\bullet}\over{1}} \bullet \partial_{\tilde{x}} g\big)  + \partial_y \big( g^{- \buildrel{\bullet}\over{1}} \bullet \partial_{\tilde{y}} g \big) = 0,  \eqno(4.15) $$ 
where $g = g(x,y,\tilde{x},\tilde{y},p,q;\hbar) \in {\cal E}(\tilde{\cal W}_D).$ This equation will be called {\it the Moyal 
deformation of Yang's equation}.

It is very interesting to note that for K\"ahler and hyper-K\"ahler 
manifolds, Yang equation admits the very natural extension [31,32]

$$ \omega \wedge \partial \big( g^{-1} \tilde{\partial} g \big) = 0, 
\eqno(4.16)$$
where $ \partial := dx \partial_x + dy \partial_y$, $\tilde{\partial}:= d\tilde{x} \partial_{\tilde{x}} + d\tilde{y} \partial_{\tilde{y}}$ and $\omega$ is here the K\"ahler two-form on the spacetime manifold 
${\bf R}^4$ where the theory is defined. This equation is well known as the {\it Donaldson-Nair-Schiff (DNS) equation}\foot{ The DNS 
equation represents a coupling between the SDYM fields and the SDG. The 
natural framework to englobe this equation seems to be N=2 heterotic 
string theory [33]. Connections to 4d analogues of WZW theory and 
Conformal Field Theories, can be found at [34].}.

The DNS equation can be derived from the so called DNS-action [31,32]

$$ S_{DNS}[g] =  { 1 \over 2} \int_{X} \omega \wedge {\rm Tr} 
\big( g^{-1} \partial g \wedge g^{-1} \tilde{\partial}g \big) -  {1 \over 3} \int_{\tilde{X}} \omega \wedge  {\rm Tr} \big( g^{-1} d g\big)^3, 
\eqno(4.17)$$
where $\tilde{X} = X \times I$ and `Tr' is an invariant form on 
su$(2)$. The above action is of 
the WZW type and it can be obtained by dimensional reduction from the 
K\"ahler-Chern-Simons theory [32].

Applying now, our method described in Section 2 we find the {\it 
Moyal deformation of DNS} 

$$ \omega \buildrel{\bullet}\over{\wedge} {\partial} \big( g^{- 
\buildrel{\bullet}\over{1}} \bullet \tilde{\partial} g \big) = 0. \eqno(4.18) $$
After a simple calculations we find that the Lagrangian from which we can derive
Eq. (4.20) is 

$$ S^{(M)}_{DNS}[g] = -{ \hbar^2 \over 2} \int_X \omega 
\buildrel{\bullet}\over{\wedge} tr \big( 
g^{ - \buildrel{\bullet}\over{1}} \bullet \partial g\buildrel{\bullet}\over{\wedge}  g^{- \buildrel{\bullet}\over{1}} 
\bullet \bar{\partial}g \big) + {\hbar^2 \over 3} 
\int_{\tilde{X}} \omega \buildrel{\bullet}\over{\wedge} tr \big( g^{-\buildrel{\bullet}\over{1}} 
\bullet d g\big)^3.\eqno(4.19)$$

\vskip 1truecm

\section{ Geometry of WZW-like Action of Self-dual Gravity}

Now we study the geometry and topology associated to the
WZW-like action obtained in Ref. [23]. We first study the case where the
basic field $\theta$ is a Lie algebra ${\cal G}$-valued function on
the spacetime manifold $\Omega$ ($ dim \ \Omega =2).$ The action reads 

$$ S (\theta) = - {\alpha\over 2} \int_{\Omega} {\rm Tr} [ d\theta 
\wedge d \theta ] + {\alpha \over 3} \int_B {\rm Tr}
(d \theta \wedge d \theta \wedge d \theta) \eqno(4.20)$$ where `Tr' is an
invariant form on the Lie algebra ${\cal G}$ of $G$ and $\theta \in 
\Lambda^0(S^2) \otimes {\cal G}$ and $\Omega$ is the boundary of $B$. The field $\theta$ can be seen as $ \theta: \Omega \to {\cal G}.$

Taking $\Omega = S^2$ one can extend this map from $S^2$ to the
3-manifold $\tilde{\Omega} = \Omega \times I$ with $\partial 
\tilde{\Omega} = 
S^2$. This is due to the fact that $\pi_2 ({\cal G})= 0.$

The maps $\overline \theta: \tilde{\Omega} \to {\cal G}$ are classified by 
$\pi_3({\cal G}).$  The triviality of this group implies that the constant
$\alpha$ must take ${\bf R}$-values i.e. $\alpha \in {\bf R}$. Thus we choose $\alpha = 1.$ Therefore the $W Z$-like
term is not in essential topological! Thus one can define globally an 
invariant 3-form $\rho$ 

$$ \rho = {1\over 3} Tr (d \theta \wedge d \theta \wedge d \theta). 
\eqno(4.21)$$

The form $\rho$ can be written globally as  an exact form $ \rho =  d 
\lambda$, where $\lambda$ is a two-form on ${\cal
G}$. Thus this term does not contribute to the classical equations of
motion. 

Action (4.22) can be generalized to Fedosov's geometry as follows

$$ S^{\bullet}(\Phi) = + {1 \over 2} \int_{S^2} tr \big[ d\Theta 
\buildrel{\bullet}\over{\wedge} d \Theta \big] - {1 \over 3} 
\int_{\tilde{\Omega}} tr  \big( d\bar{\Theta}\buildrel{\bullet}\over{\wedge} 
d\bar{\Theta} \buildrel{\bullet}\over{\wedge} d\bar{\Theta} \big). \eqno(4.22)$$
where now $\Theta \in {\cal E}(\tilde{\cal W}_D) \otimes C^{\infty}(S^2)$
and $\bar{\Theta} \in {\cal E}(\tilde{\cal W}_D)\otimes C^{\infty}(\tilde{\Omega}).$

Moyal deformation of WZW-like action for SDG can ce put in the form 
$$ S(\Phi) = +{1 \over 2} \int_{S^2} tr \big[ d\Theta 
\buildrel{\bullet}\over{\wedge} d \Theta \big] + \Gamma_{WZ}, \eqno(4.25)$$
where $ \Gamma_{WZ} \cong - \int_{\tilde{\Omega}} \bar{\Theta}^* \rho. $

\vskip 2truecm


\chapter{\bf Final Remarks}

In this paper we have reformulated different aspects of integrable
deformation of SDG in terms of a Fedosov's geometry of deformation
quantization. We find that this non-commutative geometry, appears to be a natural language to describe the SDYM and Chiral Model approaches  to SDG.

Some further questions remain to be overcome. For instance, it would be 
very interesting to investigate the behavior of heavenly hierarchies of 
conserved quantities [35,36], within Fedosov's geometry and some 
other alternative geometries of deformation quantization. Some 
applications to hyperheavenly equations would be of capital importance. 
In connection with quantum groups and non-commutative geometry might
be interesting to consider both, simultaneusly, the Moyal deformation 
and $q$-deformation. Some results of [27,37,38] might serve as begining 
point.

On the other hand, the strong connection between SDYM theory and SDG with
N=2 heterotic strings [33] might indicate that the results obtained in this
paper for SDYM (via Yang's equation), concerning its Moyal deformation
(and the geometrical interpretation) can be extended to N=2 heterotic
strings. It is very possible that we can find something like an integrable
Moyal deformation of N=2 heterotic strings. 

The surprising relation between N=2 Heterotic Strings with M and F
theories [39,40,41] and the application of the former in the searching for for 
geometrical structures of M and F theories, one would to hope that some
new geometrical descriptions of SDYM theory and SDG will be of some 
importance to give more deep insight into string theory dualities.

\vskip 2truecm

\centerline{\Ilis Acknowledgements}

We are grateful to Maciej Dunajski for send us his paper. We would like to
thank CONACyT and Academia de la Investigaci\'on Cient\'{\i}fica (AIC)
M\'exico, for partial support. One of us (M.P.) thank the staff of
Department of Physics at CINVESTAV , M\'ex ico D.F., for warm hospitality. 

\vfill
\break

\centerline{\Ilis References}

\item{[1]} I.A.B. Strachan, ``A Geometry for Multidimensional Integrable 
Systems'', J. Geom. Phys., to appear.

\item{[2]} R.S. Ward, Nucl. Phys. B {\bf 236} (1984) 381; Phil. Trans. R. 
Soc. Lond. {\bf A315} (1985) 451.

\item{[3]} L.J. Mason and G.A.J. Sparling, Phys. Lett. {\bf A137} (1989) 29.

\item{[4]} S. Chakravarty, M.J. Ablowitz and P.A. Clarkson, Phys. Rev. 
Lett. {\bf 65} (1989) 1085.

\item{[5]} L.J. Mason and E.T. Newmann, Commun. Math. Phys. {\bf 121} 
(1989) 659.

\item{[6]} R.S. Ward, Class. Quantum Grav. {\bf 7} (1990) L217.

\item{[7]} Q-Han Park, {\it Int. J. Mod. Phys.} {\bf A7}, 1415 (1992).

\item{[8]} I. Bakas, Self-duality, integrable systems, W-algebras and all 
that, in: Non-linear fields: classical random, semiclassical, eds. P.A. 
Garbaczewski and Z. Popowicz (World Scientific, Singapore, 1991).

\item{[9]} V. Husain, Phys. Rev. Lett. 72 (1994) 800; Class. Quantum 
Grav. 11 (1994) 927.

\item{[10]} J.F. Pleba\'nski, M. Przanowski and H. Garc\'{\i}a-Compe\'an, 
Acta Phys.  Pol. {\bf B25} (1994) 1079.

\item{[11]} E. Witten, J. Geom. Phys. {\bf 8} (1992) 327.

\item{[12]} R,S. Ward, Phys. Lett. {\bf B 234} (1990) 81; J. Geom. Phys. 
{\bf 8} (1992) 317.

\item{[13]} I.A.B. Strachan, J. Math. Phys. {\bf 36} (1995) 3566.

\item{[14]} I.A.B. Strachan, {\it Phys. Lett.} {\bf B283}, 63 (1992).

\item{[15]} K. Takasaki,  J. Geom. Phys. {\bf 14} (1994) 111.

\item{[16]} K. Takasaki, J. Geom. Phys. {\bf 14} (1994) 332.

\item{[17]} J.F. Pleba\'nski, M. Przanowski, B. Rajca and J. Tosiek, 
Acta. Phys. Pol. {\bf B 26} (1995) 889.

\item{[18]} J.F. Pleba\'nski, M. Przanowski and J. Tosiek, 
``Weyl-Wigner-Moyal-Formalism II. The Moyal Bracket, to appear in Acta Phys. 
Pol. {\bf B} (1996). 

\item{[19]} J.F. Pleba\'nski and M. Przanowski, ''The Universal 
Covering of Heavenly Equations Via Weyl-Wigner-Moyal Formalism'', in {\it 
Gravitation, Electromagnetism and Geometrical Structures} For the 80th 
birthday of A. Lichnerowitz, Ed. Giorgio Ferrarese, Pitagora Editrice 
Bologna (1996).

\item{[20]} J.F. Pleba\'nski and M. Przanowski, ``The Lagrangian of a 
self-dual gravitational field as a limit of the SDYM Lagrangian'', Phys. 
Lett. {\bf A 212} (1996) 22-28.

\item{[21]} J.F. Pleba\'nski M. Przanowski and H. Garc\'{\i}a-Compe\'an,
``From Principal Chiral Model to Self-dual Gravity'', Mod. Phys. Lett. {\bf
A11} (1996) 663.

\item{[22]} K. Uhlenbeck, J. Diff. Geom. {\bf 30} (1989) 1; R.S. Ward, 
Commun. Math. Phys. {\bf 123} (1990) 319.

\item{[23]} H. Garc\'{\i}a-Compe\'an, J. F. Pleba\'nski and M.. Przanowski, 
``Further remarks on the chiral model approach to self-dual gravity'', 
Phys. Lett. A {\bf 219} (1996) 249.

\item{[24]} Boris V. Fedosov, ``A Simple Geometrical Construction of 
Deformation Quantization'', J. Diff. Geom. {\bf 40} (1994) 213-238.

\item{[25]} M. de Wilde and P.B.A. Le Comite, Lett. Math. Phys. {\bf 7} 
(1983) 487-496.

\item{[26]} E. Gozzi and M. Reuter, ``Quantum-Deformed Geometry on 
Phase-Space'', DESY 92-191 UTS-DFT-92-33.

\item{[27]} M. Reuter, ``Non-Commutative Geometry on Quantum Phase-Space'', 
DESY 95-112.

\item{[28]} H. Garc\'{\i}a-Compe\'an and T. Matos, Phys. Rev. D {\bf 52} 
(1995) 4425.

\item{[29]} C.P. Boyer, J.D. Finley and J.F. Pleba\'nski, in: General 
relativity and gravitation, Einstein memorial volume, Vol. 2, ed. A. Held 
(Plenum, New York, 1980) pp. 241-281.

\item{[30]} C.N. Yang, Phys. Rev. Lett. {\bf 38} (1977) 1377.

\item{[31]} S.K. Donaldson, Proc. Lond. Math. Soc. {\bf 50} (1985) 1.

\item{[32]} V.P. Nair and J. Schiff, Nucl. Phys. B {\bf 371} (1992) 329; 
Phys. Lett. {\bf B246} (1990) 423.

\item{[33]} H. Ooguri and C. Vafa, Mod. Phys. Lett. {\bf A5} (1990) 83; 
Nucl. Phys. B {\bf 361} (1991) 469; Nucl. Phys. B {\bf 367} (1991) 83; 
Nucl. Phys.  B {\bf 451} (1995) 121.

\item{[34]} A Losev, G. Moore, N. Nekrasov and S. Shatashvili, Nucl. 
Phys. (Procc. Suppl.) {\bf 46} (1996) 130.

\item{[35]} C.P. Boyer and J.F. Pleba\'nski, J. Math. Phys. {\bf 18}
(1977) 229.

\item{[36]} M. Dunajski and J.L. Mason, ``Heavenly Hierarchies and Curved 
Twistor Spaces'', Twistor Newletter {\bf 41} (1996) August.

\item{[37]} J.F. Pleba\'nski and H. Garc\'{\i}a-Compe\'an, Int. J. Mod. 
Phys. A {\bf 10} (1995) 3371.

\item{[38]} \"O.F. Dayi, ``$q$-Deformed Star Products and Moyal 
Brackets'', Preprint q-alg/9609023, MRC.PH.TH.-12/96.

\item{[39]} E. Martinec, ``Geometrical Structures of M-Theory'', 
Preprint, hep-th/9608017.

\item{[40]} S.V. Ketov, ``From N=2 Strings to F and M Theories'', 
Preprint, hep-th/9606142; ``The $Osp(32|1)$ Versus $Osp(8|2)$ 
Supersymmetric M-brane Actions from Self-dual (2,2) Strings'', Preprint, 
hep-th/9609004; ``Mixed (open/closed) N=(2,2) String Theory as an Integrable Deformation of self-duality, hep-th/9612170.

\item{[41]} A. Jevicki, ``Matrix Models, Open Strings and Quantization of
Membranes'', hep-th/9607187.

\endpage
\end